\documentclass[twocolumn,showpacs,preprintnumbers,amsmath,amssymb, prb]{revtex4-1}

\usepackage{stmaryrd}
\usepackage{txfonts}
\usepackage{amssymb}
\usepackage{mathrsfs}
\usepackage{graphicx}
\usepackage{dcolumn}
\usepackage{bm}
\usepackage{epsfig}
\usepackage{hyperref}                 

\begin{document}
\preprint{\href{http://dx.doi.org/10.1103/PhysRevB.86.064523}{S. Z. Lin and L. N. Bulaevskii, Phys. Ref. {\bf{B}} {\bf 86}, 064523 (2012).}}

\title{Enhancement of critical current density in superconducting/magnetic multi-layers with slow magnetic relaxation dynamics and large magnetic susceptibility}

\author{Shi-Zeng Lin and Lev N. Bulaevskii}
\affiliation{Theoretical Division, Los Alamos National Laboratory, Los Alamos, New Mexico 87545, USA}

\begin{abstract}
We propose to use superconductor-magnet multi-layer structure to achieve high critical current density by invoking polaronic mechanism of pinning. The magnetic layers should have large magnetic susceptibility to enhance the coupling between vortices and magnetization in magnetic layers. The relaxation of the magnetization should be slow. When the velocity of vortices is low, they are dressed by nonuniform magnetization and move as polarons. In this case, the viscosity of vortices proportional to the magnetic relaxation time is enhanced significantly. As velocity increases, the polarons dissociate and the viscosity drops to the usual Bardeen-Stephen one, resulting in a jump in the I-V curve. Experimentally the jump shows up as a depinning transition and the corresponding current at the jump is the depinning current. For Nb and proper magnet multi-layer structure, we estimate the critical current density $J_c\sim 10^{9}\ \rm{A/m^2}$ at magnetic field $B\approx 1$ T. 
 \end{abstract}
\date{\today}
\maketitle

\section{Introduction}
One fascinating property of superconductors is ability to carry dissipationless current. With a transport current, vortices are induced inside the superconductor due to the magnetic field generated by the current. These vortices are driven by the Lorentz force exerted by the current and their motion causes voltage and dissipation. In inhomogeneous superconductors, the Lorentz force can be balanced by the pinning force due to defects. The strength of pinning thus determines how much dissipationless current the superconductor can carry, which is defined as the critical current. The random distributed point-like defects where superconductivity are weakened, are important for pinning of vortices.\cite{Blatter94} One may also introduce artificially columnar pinning centers by heavy ion irradiation.\cite{Civale91} In these cases, the pinning is caused by suppression of superconductivity.

An alternative approach to introduce pinning is to use magnetic moments, which interact strongly with vortex. Such option may be present in magnetic superconductors. \cite{Bulaevskii85,Buzdin05} The magnetic moments can also be introduced artificially in hybrid systems consisting of superconducting and magnetic layers\cite{Lyuksyutov05}. It was proposed in Ref. \onlinecite{Bulaevskii00} that the domain walls can provide strong pinning with characteristic pinning energy $\Phi_0 M_d$. Here $\Phi_0=hc/(2e)$ is the quantum flux and $M_d$ is the magnetization at the wall. There are experimental attempts to enhance the critical current by putting magnetic particles\cite{Otani1993}, dots\cite{Martin97,Martin99} or ferromagnet with domain walls on top of superconductors. \cite{Vlasko08}

Reduction of dissipation can be also achieved by enhancement of the vortex viscosity. At a given current $J$, the dissipation power for a superconductor without pinning due to quenched disorder is proportional to $J^2/\eta$ where $\eta$ is the vortex viscosity. In nonmagnetic superconductors, $\eta$ is just the standard Bardeen-Stephen (BS) drag coefficient accounting for the dissipation in the normal vortex core. If one can increase significantly the vortex viscosity, superconductors can carry large current density with low dissipation, despite vortices are not pinned. It was shown that in magnetic superconductors, motion of vortex lattice excites magnons\cite{Shekhter11}. When the kinematic condition $\Omega(\mathbf{G})=\mathbf{G}\cdot\mathbf{v}$ is satisfied, Cherenkov radiation of magnon occurs and the vortex viscosity is enhanced due to transferring energy into the magnetic subsystem, where energy is finally dissipated through magnetic damping. Here $\mathbf{G}$ is the lattice wave vector, $\mathbf{v}$ is the velocity of vortex lattice and $\Omega(\mathbf{G})$ is the magnon spectrum.  When the magnetic damping is weak, magnetic domain walls are created dynamically due to the parametric instability and the viscosity is increased further.\cite{szlin12a}

Recently a polaronic mechanism of vortex pinning is proposed in Ref. \onlinecite{Bulaevskii12a} to explain the increase of critical current observed in ErNi$_2$B$_2$C below the incommensurate to commensurate spin density wave (SDW) transition at 2.3 K\cite{Gammel2000}. The transition into the commensurate SDW phase leaves 1/20 spins free from molecular field\cite{Canfield1996}. These spins can be easily polarized by vortices.  These spins are Ising spins and experience large crystal field splitting\cite{Gasser1997}, which results in slow relaxation dynamics\cite{Bonville1996}. When the velocity of vortex lattice is low, $a/v\gg \tau$, the nonuniform component of free-spin magnetization induced by vortex lattice follows the vortex motion, and the nonuniform magnetization and vortex form a polaron. Here $a$ is the vortex lattice constant and $\tau$ is the relaxation time for magnetizations. The effective viscosity of vortex lattice increases with the relaxation time. For a large velocity, $a/v\ll \tau$, the nonuniform magnetization cannot follow the motion of vortex lattice and they are decoupled from each other. The viscosity of the system recovers to the conventional BS one. The decoupling or dissociation of polaron experimentally shows up as a depinning transition. The maximal critical current for ErNi$_2$B$_2$C is estimated as $10^{10}\ \rm{A/m^2}$ at magnetic field $B\approx 0.1$ T. The polaronic mechanism is also at work in other borocarbides, cuprate and iron-based superconductors with magnetic rear earth ions locating between superconducting layers.   

\begin{figure}[t]
\psfig{figure=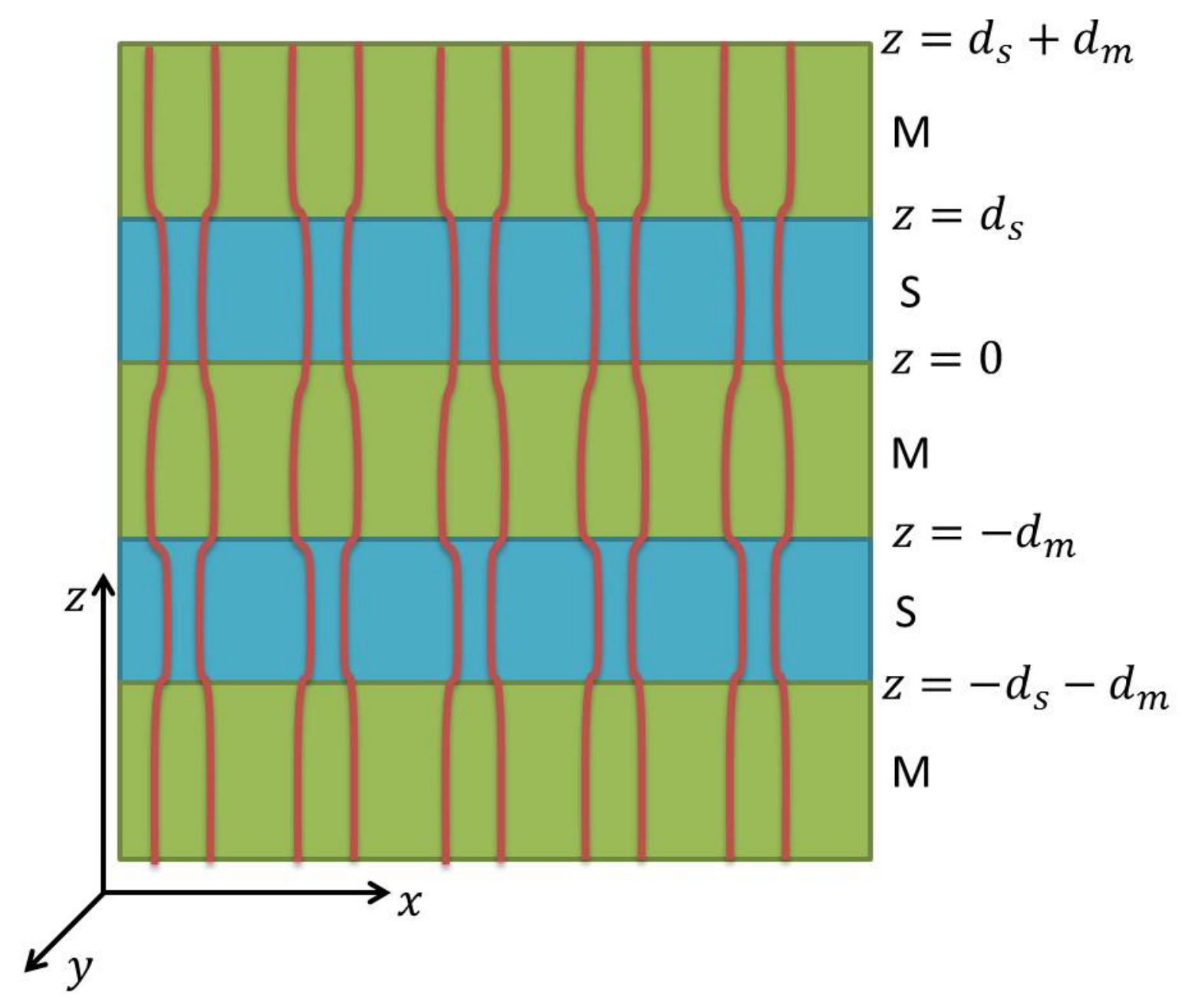,width=\columnwidth}
\caption{\label{f1}(color online) Schematic view of multi-layer structure consisting of alternating magnetic (M) layers (green) with thickness $d_m$ and superconducting (S) layers (blue) with thickness $d_s$. The distribution of the magnetic field is shown by red lines. } 
\end{figure}

The polaronic mechanism of pinning provides an additional routine to achieve high critical current. To optimize such pinning mechanism, we propose to use a multi-layer structure consisting of superconducting (S) and magnetic (M) layers shown in Fig. \ref{f1}, to achieve high critical current. For that the magnetic layers should have high magnetic susceptibility at working magnetic field to ensure a strong coupling between magnetic moments and vortices. Secondly, the relaxation time of the magnetization should be long. Thirdly, the penetration depth of the superconducting layers should be small.

\section{Model and results}
 
Under external magnetic fields, the vortex lattice is induced inside the S layers. With a transport current, vortex lattice moves in response to the Lorentz force. In the quasistatic approximation, the motion of vortex lattice is given by
\begin{equation}\label{eq1}
\lambda^2\nabla \times \nabla \times \mathbf{B}+\mathbf{B}=\Phi _0\sum_i\delta \left[\mathbf{r}-\mathbf{r}_i(t)\right]\hat{\mathbf{z}},
\end{equation}
where $\mathbf{r}_i(t)=\mathbf{r}_0-\mathbf{v} t$ is the vortex coordinate, $\hat{\mathbf{z}}$ is the unit vector along the $z$ axis and $\lambda$ is the London penetration depth. In the flux flow region, the quenched disorder is averaged out by vortex motion and the lattice ordering is improved\cite{Koshelev94,Besseling03}. The magnetic field inside the M layers is determined by the Maxwell equations
\begin{equation}
\nabla\times (\mathbf{B}-4\pi\mathbf{M})=0, \ \ \nabla\cdot \mathbf{B}=0.
\end{equation}
The magnetization $\mathbf{M}$ depends on $\mathbf{B}$ and is determined by the material properties.  With a strong field and in static case, $\mathbf{M}$ is a nonlinear function of $\mathbf{B}$ and generally can be expressed as $\mathbf{M}(\mathbf{r})=\int d\mathbf{r}^3 f(\mathbf{r}-\mathbf{r}', \mathbf{B(r')})$.  The characteristic length of magnetic subsystem is much smaller than $\lambda$ and we use a local approximation $f(\mathbf{r}-\mathbf{r}', \mathbf{B(r')})=\delta(\mathbf{r}-\mathbf{r}')f(\mathbf{B(r')})$. $\mathbf{B}(\mathbf{r})$ has component uniform in space, ${\mathbf{B}_0}$, and the other component nonuniform in space, $\tilde{\mathbf{B}}(\mathbf{r})$, with ${\mathbf{B}_0}\ll \bar{\mathbf{B}}$. Thus the spatially nonuniform magnetization $\tilde{\mathbf{M}}(\mathbf{r})$ is $\tilde{\mathbf{M}}(\mathbf{r})\approx \partial f(\mathbf{B}_0)/\partial B_0 \tilde{\mathbf{B}}(\mathbf{r})\equiv \chi_0 (\mathbf{B}_0) \tilde{\mathbf{B}}(\mathbf{r}) $. In the following we assume the magnetic subsystem is isotropic and is characterized by a susceptibility $\chi_0(\mathbf{B}_0)$ at $\mathbf{B}_0$ in static case. The magnetic field inside the M layer is determined by the equation $\nabla^2\tilde{\mathbf{B}}=0$. Since only the spatially nonuniform component $\tilde{\mathbf{M}}$ and $\tilde{\mathbf{B}}$ are responsible for pinning, we will focus on the nonuniform components in the following calculations. At the interface between the M and S layers, we use the standard boundary condition for the field parallel to the $z$-axis $B^z$ and field parallel to the interface $B^{||}$ 
\begin{equation}\label{eq2}
B^z|_{\rm{S}}=B^z|_{\rm{M}},\ \ \ \ B^{||}|_{\rm{S}}=(1-4\pi\chi_0)B^{||}|_{\rm{M}}.
\end{equation}
Then we can obtain the magnetic field inside the M layers
\begin{equation}\label{eq3}
B_m^z(G>0, z)=\alpha\left[e^{G z'}+e^{- G(z'+d_m)}\right]\frac{\Phi _0\exp \left(-i G_x v_x t\right)}{1+\lambda ^2G^2},
\end{equation}
\begin{equation}\label{eq4}
B_m^{||}(G>0, z)=i \alpha\left[e^{G z'}-e^{- G(z'+d_m)}\right]\frac{\Phi _0\exp \left(-i G_x v_x t\right)}{1+\lambda ^2G^2},
\end{equation}
\[
\alpha=-\frac{ e^{ d_m G} \left(-1+e^{d_s k_s}\right) \chi '}{(1-\chi')(e^{d_s k_s}-e^{G d_m})+(1+\chi')(1-e^{d_m G+d_s k_s})},
\]
with $z'=z-n(d_s+d_m)$, $\ k_s=\sqrt{\lambda^{-2}+G^2}$, and $\chi'=(1-4\pi\chi)^{-1}k_s/G$.
Here $n$ is the layer index and the vortex motion is assumed to be along the $x$ direction. We consider square lattice $\mathbf{G}=(m_x2\pi/a, m_y 2\pi/a)$ with $a=\sqrt{\Phi_0/B_0}$ the lattice constant and $m_x$, $m_y$ integers. 

We assume a relaxational dynamics for the M layers, $\mathbf{M}(\omega)=\chi(\omega) \mathbf{B}_m(\omega)$,  with a dynamic susceptibility 
\begin{equation}\label{eq5}
\chi(\omega)=\frac{\chi_0}{1+i\omega\tau}.
\end{equation}
Here we have assumed that the relaxation dynamics is governed by a single relaxation time. This assumption is not essential but just for convenience of calculations. In the steady state, we have 
\begin{equation}\label{eq6}
\mathbf{M}\left(\mathbf{G}, z, t\right)=\int _0^t\exp [(t'-t)/\tau ]\frac{\chi _0\mathbf{B}_m(\mathbf{G}, z, t')}{\tau }dt'.
\end{equation}
Because of the relaxation, $\mathbf{M}$ depends on the history of vortex motion. Due to slow relaxation of the magnetization, there is retardation between the time variation of induced nonuniform magnetization and vortex motion. As a result, the magnetization exerts a drag force to the vortex which is opposite to the driving force. The pinning force acting on a single vortex due to the induced magnetization in one M layer is given by $F_p=\partial_{r_0}\int dxdy\int_{-d_m}^0 dz\mathbf{M}\cdot\mathbf{B}_m$, which yields
\begin{equation}\label{eq7}
F_p=\sum_G \left[1-\exp \left(-2G d_m\right)\right]\frac{2\alpha^2\chi _0}{\left(1+\lambda ^2G^2\right)^2 a^2}\frac{ G v \tau \Phi _0^2}{1+ (G v \tau)^2 }.
\end{equation}
The I-V curve is determined by the equation of motion for vortex $d_s\eta_{BS} v=d_s F_L -F_p$ with the electric field $E=B v/c$ and the Lorentz force $F_L=J\Phi_0/c$. Here $\eta_{BS}$ is the BS viscosity $\eta_{BS}=\Phi_0^2/(2\pi\xi^2 c^2\rho_n)$ with $\rho_n$ the resistivity just above $T_c$ and $\xi$ the coherence length. We consider a realistic case where $a/(2\pi)\ll d_m, d_s$. Taking into account only the dominant contribution $G_x=2\pi/a$ and $G_y=0$ in the summation, we obtain
\begin{equation}\label{eq8}
u=\mathcal{F}_L-\mathcal{F}_p\frac{ u}{1+ u^2},
\end{equation}
\begin{equation}\label{eq8aa}
\mathcal{F}_L=\frac{F_L}{\eta_{BS}  v_0 },\ \  \mathcal{F}_p=\frac{2\tau}{\eta_{BS} d_s}\left(\frac{1}{2-4\pi  \chi _0}\right)^2\frac{\chi _0 a\Phi _0^2}{\lambda ^4(2\pi)^3},
\end{equation}
with $u=v/v_0$ and $v_0=a/(2\pi\tau)$.

\begin{figure}[t]
\psfig{figure=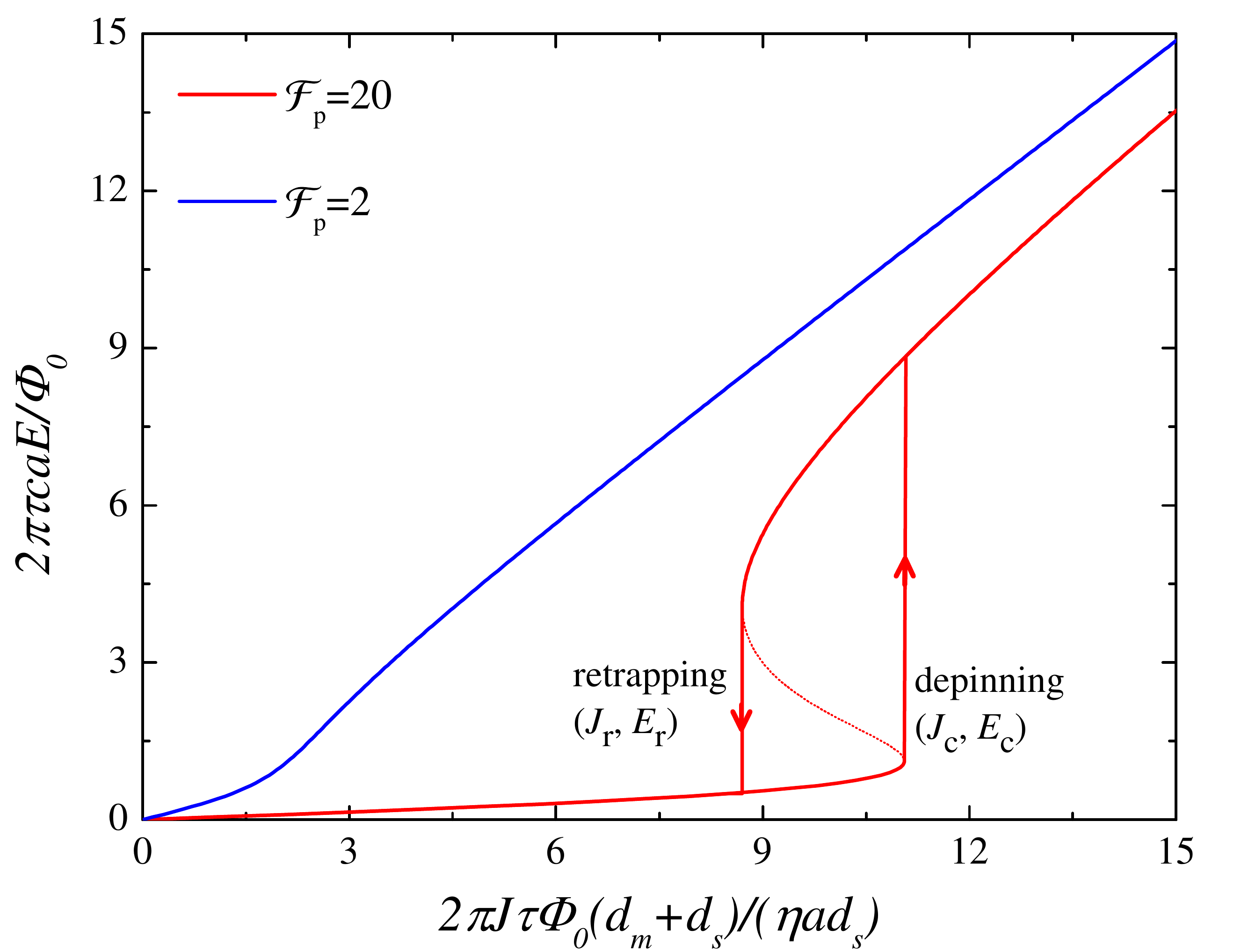,width=\columnwidth}
\caption{\label{f2}(color online) Calculated I-V curves for $\mathcal{F}_p=20$ and $\mathcal{F}_p=2$. For $\mathcal{F}_p=20$ the system shows hysteresis in the I-V while for $\mathcal{F}_p=2$ no hysteresis is present. The green dotted line denotes the unstable solution.} 
\end{figure}

At a small velocity $u\ll 1$, the velocity is given by $u=\mathcal{F}_L/(1+\mathcal{F}_p)$ which becomes inversely proportional to $\tau$ for a large $\tau$. For a large $u\gg 1$, we recover the conventional BS viscosity $v=\mathcal{F}_L$. The dependence of $u$ on $\mathcal{F}_L$ is shown in Fig. \ref{f2}. Hysteresis is developped when $\mathcal{F}_p\ge 8$. For typical parameters for Nb superconductor $\xi\approx\lambda\approx 40$ nm, $\rho_n \approx 10^{-6}\ \rm{\Omega\cdot m}$ and $a=40$ nm at $B\approx 1$ T and $\chi_0=0.05$, $\mathcal{F}_p\ge 8$ requires $\tau>1$ ps. For the relaxation time of order $\tau\approx 1\ \rm{\mu s}$, the effective viscosity is enhanced by a factor of $10^{6}$ compared to the bare BS one at $v< a/\tau$. Upon increasing the current, the velocity increases and at a critical current (velocity), the system jumps at $d\mathcal{F}_L/d u=0$ to the conventional BS branch due to the dissociation of vortex polaron. The effective critical current density for the whole system is given by
\begin{equation}\label{eq9}
J_c\approx\left(\frac{1}{2-4\pi  \chi _0}\right)^2\frac{\chi _0c}{(2\pi )^4\lambda ^4}\frac{\Phi_0 a^2}{d_s+d_m}.  
\end{equation}
For $d_s=d_m=100$ nm , we obtain $J_c\approx 10^9\ \rm{A/m^2}$. On the other hand, when one reduces the current from the conventional BS branch, the system jumps to the branches with high viscosity due to the formation of vortex polaron. We call this retrapping transition and the retrapping current $J_r$ is
\begin{equation}\label{eq10}
J_r\approx \frac{1}{1-2\pi  \chi _0}\sqrt{\frac{\chi_0\eta_{BS} a d_s}{\pi \tau}}\frac{a c}{\lambda^2 4\pi^2} \frac{1}{d_s+d_m}.
\end{equation}
For the parameters used before and $\tau=1\ \rm{{\mu s}}$, we estimate $J_r\approx 2\times 10^6\ \rm{A/m^2}$.

\section{Discussions}

Now we discuss the optimal thickness of M and S layers. For $d_m\gg a$, the magnetic induction and the magnetization is practically uniform in the lateral direction in the middle of the M layer. This can be seen from Eqs. (\ref{eq3}) and (\ref{eq4}) that $B_m(G>0)\approx \exp(-2\pi d_m/a)$ when $-d_m\ll z'\ll 0$. As a result, the pinning force becomes practically $d_m$ independent when $d_m\gg a$. In other words, the pinning is effective only near the boundaries between S and M layers in the area of thickness of the order $a$. On the other hand, the Lorentz force is proportional $d_s$. Thus the effective critical current of the whole system $J_c$ is proportional to $1/(d_s+d_m)$ as described by Eq. (\ref{eq9}). Therefore the thinner of both M and S layers, the higher is the critical current of the system.  
 
Let us discuss the possible choice of S and M layers. The critical current decreases as $\lambda^{-4}$ because the smaller $\lambda$, the more nonuniform is the magnetic field distribution inside the M layers, hence stronger pinning. Thus superconductors with smaller $\lambda$ are preferred. The critical current does not depend on $\tau$ for sufficiently large $\tau$, while the viscosity in the branch with vortex polaron is proportional to $\tau$. The slow magnetic dynamics can be realized in spin glasses. Their relaxation is described by  a broad spectrum of time scale, with average time of the order $0.1\ \rm{\mu s}$\cite{Uemura1985,Binder1986}. For $\rm{CuMn_{0.08}}$, $\chi_0\approx 0.002$ at $B=1$ T. \cite{Prejean1980} One may enhance $\chi_0$ by tuning the concentration of magnetic metal in alloys. \cite{Kouvel1961} One may use superparamagnets with $\tau$ as large as $1$ s and with huge $\chi_0$ due to large magnetic moments in superparamagnets. \cite{Bean1959,Goldfarb1981,Cullity2008} One may also use the recently synthesized cobalt-based and rare-earth-based single chain magnets with $\chi_0\approx 0.05$ at $B=1$ T and $10^{-6}\ \rm{s}<\tau<10^{-4}\ \rm{s}$.\cite{Caneschi2001,Caneschi2002,Bogani2005,Bernot2006}.

Next we discuss the effect of quenched disorder. In the presence of quenched disorder, the vortex lines adjust themselves to take the advantage of the pinning potential, which destroys the long-range lattice order. Below a threshold current, vortices remain pinned (actually they creep between pinning centers due to fluctuations). In this region, the polaronic mechanism does not play a role. When the current is high enough to depin the vortices from quenched disorder, vortices start to move and the lattice ordering is enhanced. By formation of polaron with the nonuniformly induced magnetization, the vortex viscosity is enhanced. At a critical velocity (current), the polaron dissociates and the system jumps to the conventional BS branch. Pinning due to quenched disorder works in the static region and polaronic pinning works in the dynamic region. The critical current of the whole system is the sum of these two threshold currents. Note that magnetostriction in combination with quenched disorder enhance the polaronic pinning mechanism.

The M/S multi-layer structure is naturally in some superconducting single crystals, such as  $\rm{RuSr_2GdCu_2O_8}$\cite{McLaughlin99} and (RE)Ba$_2$Cu$_3$O$_7$\cite{Hor1987,Allenspach1995}, where RE is the rear earth magnetic ions. In  $\rm{RuSr_2GdCu_2O_8}$ the magnetic moments order ferromagnetically above $T_c$ thus the dominant enhancement of vortex viscosity is due to the radiation of magnons\cite{Shekhter11}. For (RE)Ba$_2$Cu$_3$O$_7$, magnetic RE ions positioned between superconducting layers interact weakly with superconducting electrons and order at very low N\'{e}el temperatures of the order $T_N\sim 1$ K. The polaronic mechanism is important above the magnetic ordering temperature, where spins are free. The London penetration depth of cuprate superconductors is large $\lambda\approx 200$ nm, thus the critical current is reduced significantly compared to that for Nb multi-layer structure, because $J_c$ drops as $1/\lambda^4$. Another natural realization is the recently discovered iron-based superconductors, such as $\rm{(RE)FeAsO_{1-x}F_{x}}$, where $\rm{RE}$ ions ordered antiferromagnetically below $T_N\sim 1$K.\cite{Wang2008}

\section{Conclusion}
To summarize, we have proposed superconductor-magnet multi-layer structure to achieve high critical current density based on the polaronic pinning mechanism. The critical current is estimated to be $10^{9}\ \rm{A/m^2}$ at $B\approx 1$ T for an optimal configurations of Nb and proper magnet multi-layer structure. In the presence of quenched disorder, the polaronic pinning starts to work when vortices depin from quenched potential. Thus the total critical current of the system is the sum of depinning current due to quenched disorder and depinning current due to the polaronic mechanism.

\acknowledgments
The authors are indebted to Cristian D. Batista for helpful discussion. This publication was made possible by funding from the Los Alamos Laboratory Directed Research and Development Program, project number 20110138ER.

%

\end{document}